\documentclass[a4paper,12pt,reqno]{amsart}
\usepackage{amssymb,a4,t1enc}

\newtheorem{Def}{Definition}[section]

\newcommand{\menge}[1]{\mathbb{#1}} 
 
\newcommand{\dif}{\ensuremath{\mathrm{d}}}

\newcommand{\D}{\ensuremath{\partial}}
 
\newcommand{\R}{\ensuremath{\menge{R}}} 
\newcommand{\Rd}{\ensuremath{\menge{R}^{d}}} 
 
\newcommand{\Rv}{\ensuremath{\menge{R}^{4}}}

\newcommand{\Dcal}{\ensuremath{\mathcal{D}}}

\newcommand{\scp}[2]{\ensuremath{\left\langle #1, #2 \right\rangle}}

\begin{document}

\title[Lorentz Invariant Renormalization]%
{Lorentz Invariant Renormalization in Causal Perturbation Theory}
\author{K. Bresser, G. Pinter and D. Prange}

\maketitle

\vspace{-1.1cm}
\begin{center}
\textsc{%
II. Institut f\"ur Theoretische Physik\\
Universit\"at Hamburg\\
Luruper Chaussee 149\\
22761 Hamburg 
Germany}\\
e-mail: klaus.bresser@desy.de, gudrun.pinter@desy.de,
dirk.prange@desy.de
\end{center}
 
\vspace{1cm}
\begin{abstract}
\noindent
In the framework of causal perturbation theory renormalization
consists of the extension of distributions. We give the explicit form
of a Lorentz invariant extension of a scalar distribution, depending
on one difference of space time coordinates.
\end{abstract}

\section{Introduction}

Causal perturbation theory was founded by the work of
\cite{stue}, \cite{meister} and \cite{epstein}. In this approach,
the $S$-matrix is constructed as a formal functional expansion
in a so-called switching function of the interaction. The 
coefficients in the expansion, determined by induction, are
time ordered products of operator valued distributions.

The advantage of this approach in comparison with the derivation
of the $S$-matrix in the Hamilton formalism is that the time
plays no distinguished role, and only a few physical properties
of the $S$-matrix like causality, locality and unitarity
are assumed.

The definition of the time ordered products in the $S$-matrix requires
an extension of their domain of definition which corresponds to
renormalization.  This is briefly discussed in section \ref{ab1}.  To
obtain a Lorentz invariant $S$-matrix, the extension has to be Lorentz
invariant.  The existence of such an extension was first proved in
\cite{epstein}, later it was discussed in \cite{stora} and
\cite{scharf} as a cohomological problem.

In spite of the knowledge of its existence the explicit form of this
extension has not been calculated up to now. We fill this gap, giving
a solution for a distribution in one argument in section \ref{ab2}.

In the last section we calculate the dependence of the Lorentz
invariant extension on the auxiliary function needed in the extension
procedure.

\section{From the $S$-matrix to the
Extension of Distributions} \label{ab1}
In causal perturbation theory the $S$-matrix
is expanded as a functional of the testfunction $g \in 
\mathcal{D} (\Rv)$ switching the coupling:
\begin{equation}
S \left( g \right) = 1 + \sum_{n=1} ^{\infty} \frac{i^n}
{n!} \int \dif^4 x_1 \ldots \int \dif^4 x_n 
T \left( \mathcal{L}_{int} \left( x_1 \right)
\ldots \mathcal{L}_{int}  \left( x_n \right)  \right) 
g \left( x_1 \right)
\ldots  g \left(  x_n \right).
\end{equation}
The coefficients of this expansion are the time ordered 
($T$-) products of the interaction part $ \mathcal{L}_{int}$
of the Lagrangian density. 
They are symmetric operator valued Lorentz invariant 
distributions on a dense domain of the Fock space of free fields.
A $T$-product of given order $n$ is defined by all $T$-products
of lower order up to the total diagonal $\Delta= \{ (x_1, \ldots, x_n)
| x_i = x_j \ \forall i,j \}$ through causality:
$T | \mathcal{D} \left( \R ^{4n} \setminus \Delta \right) = {^0T}$. 
Using Wick's theorem ${^0T}$
can be decomposed into a sum of products of Lorentz invariant, 
translation invariant numerical distributions and Wick ordered free
field operators,
\begin{equation}
{^0T} \left(  \mathcal{L}_{int} \left( x_1 \right)
\ldots  \mathcal{L}_{int} \left( x_n \right) \right)
= \sum _i {^0t}_i \left( x_1, \ldots, x_n \right)
: A_i \left( x_1, \ldots, x_n \right):.
\end{equation}
Note that for noncoincident points ${^0t}$ is a product of
Feynman propagators.
According to theorem 0 of \cite{epstein} it is sufficient to 
extend the numerical part ${^0t}_i$ to the diagonal to find $T$.
This is the inductive causal construction.

An extension of the numerical distribution ${^0t}$ always exists,
but the uniqueness of the extension depends on the singular
behaviour of ${^0t}$ at the total diagonal \cite{fred}.
By choosing difference coordinates, the
singularity is shifted to the origin.
The singular behaviour at the origin
is described by the scaling degree and the
singular order.

\begin{Def}
The scaling degree $\delta$ of a numerical distribution $t$ 
at $0$ is defined by
\begin{equation}
\delta := \inf \left\{ \delta ' \left| \lim_{\epsilon \to 0}
\epsilon^{\delta'} t \left( \epsilon x  \right) =0 \right. \right\}.
\end{equation}
\end{Def}
\begin{Def}
The singular order $\omega$ of a numerical distribution $t$
at $0$ is defined by
\begin{equation}
\omega = \left[ \delta - d \right] \label{sing}
\end{equation}
where $d$ is the space time dimension and $[x]$ denotes the largest
integer $n$ with $n \leq x$.
\end{Def}
With $\mathcal{D}_{\omega}  \left( \R^{4n} \right)$ we denote the
subspace of testfunctions vanishing to order $\omega$ at $0$: 
\begin{equation}
\mathcal{D}_{\omega}  \left( \R^{4n} \right) =
\left\{ \psi \in \mathcal{D} \left( \R^{4n} \right) |
D^{\alpha} \psi (0) = 0 \ \forall \ 
|\alpha| \leq \omega \right\},
\end{equation}
where $\alpha$ is a multiindex.  Distributions with singular order
$\omega$ are well defined on $\mathcal{D}_{\omega}$ \cite{fred}.  The
extension of the numerical distribution to $\mathcal{D}$ is achieved
with the help of a projection operator, acting on the testfunctions.
\begin{Def}
  To each $\omega \in \mathbb{N}_0$ and each $ w\in \Dcal
  _0^{\infty}$ fulfilling $w(0)=1$, $D^{\alpha}w(0) =0$ $\forall$
  multiindices $\alpha$ with $1 \leq |\alpha| \leq \omega$, we
  associate a projection operator
\begin{align}
  W \left( \omega, w \right): \ \mathcal{D} \left( \R^{4n} \right)
  &\rightarrow \ \mathcal{D}_{\omega} \left( \R^{4n}
  \right)  \notag \\
  \phi \left( x \right) & \mapsto \phi \left( x \right) - w (x)
  \sum_{|\alpha| \leq \omega} \frac{x^{\alpha}}{\alpha !}  D^{\alpha}
  \phi \left( 0 \right).
\end{align}
\end{Def}
$W$ is a modified Taylor subtraction operator.
Since the function $w$ has compact support, the result is a function with 
compact support and vanishes to order $\omega$ at 
$0$. Therefore the distribution ${^0t}$ with singular order $\omega$
is defined on all $ W \left( \omega , w\right) \phi $.
The extension of ${^0t}$ to $ \mathcal{D}'$
has the following form \cite{fred}:
\begin{equation}
\left\langle t,\phi \right\rangle  =  \left\langle {^0t}, W \left( 
\omega, w \right)
\phi \right\rangle 
+ \sum_{|\alpha| \leq \omega} 
\frac{c^{\alpha}}{\alpha !}
D^{\alpha} \phi (0).  \label{fort}
\end{equation}
If the singular order $\omega$ of ${^0t}$ is negative, the extension
is unique, in the other case there are free constants $c^{\alpha}$ in
the extension procedure.  The form of the $c^{\alpha}$ can be
restricted by demanding invariance of $t$ under symmetry operations,
e.g. under the action of the Lorentz group.

\section{The Lorentz-Invariant Extension in Scalar Theories}
\label{ab2}
It is possible to determine the free constants $c^{\alpha}$ of an
extension $t$ such that Lorentz invariance of ${^0t}$ is preserved
\cite{epstein, stora, scharf}. We denote the action of the Lorentz
group on $\R ^{4n}$ by $ x \to D \left( \Lambda \right) x$. This leads
naturally to an action on $\mathcal{D} \left( \R ^{4n} \right)$ and
$\mathcal{D}' \left( \R ^{4n} \right)$, respectively, given by
\begin{align}
  \mathcal{D} \left( \R ^{4n} \right) \ni \phi &\mapsto\phi_{\Lambda}, 
  & \phi _{\Lambda}(x)
  &:= \phi \left(D \left( \Lambda ^{-1} \right) x \right) \notag\\
  \mathcal{D}' \left( \R ^{4n} \right) \ni t &\mapsto t _{\Lambda}, 
  & \left\langle t _{\Lambda}, \phi \right\rangle & := \left\langle t ,
  \phi _{\Lambda ^{-1}} \right\rangle.  \notag
\end{align}
According to equation (\ref{fort}), the extension has the form
\begin{equation}
\left\langle t ,\phi \right\rangle  =  \left\langle {^0t}, \phi -  w
\sum_{|\alpha| \leq
\omega} \frac{x^{\alpha}}{\alpha !} D^{\alpha} \phi (0)\right\rangle
+ \sum_{|\alpha| \leq
\omega} \frac{c^{\alpha}}{\alpha !} D ^{\alpha} \phi (0). 
\label{gl1}
\end{equation}
After performing a Lorentz transformation we obtain
\begin{equation}
\left\langle t_{\Lambda} , \phi \right\rangle =
\left\langle t ,\phi _{\Lambda ^{-1}} \right\rangle  
=  \left\langle {^0t}, \phi  -  w _{\Lambda }
\sum_{|\alpha| \leq
\omega} \frac{x^{\alpha}}{\alpha !} D^{\alpha} \phi (0)\right\rangle
+ \sum_{|\alpha| \leq
\omega} \frac{\left( D(\Lambda)
c \right)^{\alpha}}{\alpha !} 
D ^{\alpha} \phi (0). \label{gl2}
\end{equation} 
In order to be a Lorentz invariant extension, the difference of
(\ref{gl1}) and (\ref{gl2}) has to be zero:
\begin{equation}
\sum_{|\alpha| \leq \omega} \frac{D^{\alpha} \phi (0)}{\alpha !}
\left\langle {^0t}, \left( w- w_{\Lambda } \right)
x^{\alpha}
\right\rangle = - \sum_{|\alpha| \leq \omega}
\left( \left( D(\Lambda ) 
-1 \right) c \right) ^{\alpha}
\frac{D^{\alpha} \phi (0)} {\alpha !} .
\end{equation}
So we have to solve
\begin{equation}
\left\langle {^0t}, \left( w- w_{\Lambda } 
\right)x^{\alpha}\right\rangle = - \left(
\left( D(\Lambda) -1 \right) c \right) ^{\alpha} \label{stern}
\end{equation}
for all $\alpha$, where $ c^{\alpha} $ is a tensor of rank $|\alpha|$
and $D(\Lambda)$ is the corresponding tensor representation of the
Lorentz group.

From now on, we restrict ourselves to the case of distributions in one
coordinate. In this case only the totally symmetric part of the $
c^{\alpha} $ contributes to (\ref{gl1}).  Using Lorentz indices,
(\ref{stern}) reads
\begin{equation}
\left\langle {^0t},  \left( w- w_{\Lambda} \right)x^{\mu_1}\ldots
x^{\mu_n}\right\rangle = - \left( {\Lambda ^{\mu_1}}_{\beta_1} \ldots
 {\Lambda ^{\mu_n}}_{\beta_n} - \delta  ^{\mu_1}_{\beta_1} \ldots
 \delta  ^{\mu_n}_{\beta_n} \right) c^{\beta_1 \ldots \beta_n} 
\label{stern1}
\end{equation}
where $n= |\alpha|$. Using infinitesimal transformations we can solve
these equations for $c$ inductively.

In the case $|\alpha|=0$, (\ref{stern}) is fulfilled for all choices of
$c$, since the 1-dimensional representation of the Lorentz group is
trivial.  For $|\alpha| \geq 1$, the solution is unique up to Lorentz
invariant contributions consisting only of symmetrized tensor products
of the metric tensor $g^{ \mu \nu}$ (which generate Lorentz invariant
counterterms like $ \square \delta(x)$).

We use the generators of Lorentz transformations
\begin{equation}
{\left( l^{\alpha \beta} \right)^{\nu}}_{\mu} = g^{\beta \nu}
\delta_{\mu}^{\alpha} - g^{\alpha\nu}\delta_{\mu}^{\beta}. \label{form}
\end{equation}
The representation of a Lorentz transformation on $\R ^4$ has the form
\begin{equation}
{\Lambda ^{\nu}}_{\mu} = \delta ^{\nu}_{\mu} 
+ \frac{1}{2} \Theta_{\alpha \beta}
{(l^{\alpha \beta})^{\nu}}_{\mu} + O(\Theta ^2) \label{abcd}
\end{equation}
with infinitesimal parameters $\Theta _{\alpha \beta}$ satisfying
$\Theta _{\alpha \beta}= -\Theta _{\beta \alpha}$.  We obtain
\begin{equation}
w(x) - w(\Lambda ^{-1} x) = \frac{1}{2} \Theta _{\alpha \beta}
{(l^{\alpha \beta})^{\nu}}_{\mu} x^{\mu} \partial_{\nu} w(x)
+ O (\Theta ^2). \label{abc}
\end{equation}
We use the abbreviations
\begin{equation}
n!! = \left \{ \begin{array}{r@{\quad}l}
2 \cdot 4 \cdot \ldots \cdot n & \mbox{ for $n$  even} \\
1 \cdot 3 \cdot \ldots \cdot n & \mbox{ for $n$  odd}
\end{array} \right. 
\end{equation}
and
\begin{equation}
b^{(\alpha _1 \ldots \alpha_n)}= \frac{1}{n!} \sum_{\pi \in S_n} 
b^{\alpha_{\pi (1)} \ldots \alpha_{\pi (n)}}
\end{equation}
for the total symmetric part of a tensor $b$.  Now we prove by
induction over $|\alpha|$ that the symmetric part of $c^{\alpha}$ for
$|\alpha|>0$ is given by (up to the aforementioned ambiguity for even
$|\alpha|$):
\begin{multline}
  c^{(\alpha _1 \ldots \alpha_n)} = \frac{ (n-1)!!}{(n+2)!!}
  \sum_{s=0}^{\left[\frac{n-1}{2} \right] }
  \frac{(n-2s)!!}{(n-1-2s)!!}  g^{(\alpha_1 \alpha _2} \ldots
  g^{\alpha _{2s-1} \alpha _{2s}} \times
  \\ \times
  \left\langle {^0t}, (x^2)^s x ^{\alpha _{2s+1}} \ldots x ^{\alpha
      _{n-1}} \left( x^2 \partial ^{\alpha _{n})} w - x ^{\alpha
        _{n})} x^{\beta} \partial _{\beta} w \right) \right\rangle.
\label{result}
\end{multline}

At the beginning of the induction we determine the $c^{\alpha}$ for
$|\alpha|=1$ and $|\alpha|=2$.
\begin{enumerate}
\item $|\alpha|=1$. We obtain
\begin{align}
  \left( \left( D(\Lambda) -1 \right) c \right)^{\nu} &= \frac{1}{2}
  \Theta _{\alpha \beta}
  {(l^{\alpha \beta})^{\nu}}_{\mu} c^{\mu} \notag \\
  &\stackrel{(\ref{abc})(\ref{stern})}{=} - \left\langle {^0t},
    \frac{1}{2} \Theta _{\alpha \beta} {(l^{\alpha
        \beta})^{\rho}}_{\mu} x^{\mu} \partial_{\rho} w
    x^{\nu}\right\rangle,
\end{align}
which yields (independence of $\Theta _{\alpha \beta}$)
\begin{equation}
 {(l^{\alpha \beta})^{\nu}}_{\mu} c^{\mu} = -\left\langle {^0t},
 {(l^{\alpha \beta})^{\rho}}_{\sigma} x^{\sigma} \partial_{\rho} w
x^{\nu}\right\rangle.
\end{equation}
Inserting the form (\ref{form}) of the $(l^{\alpha \beta})$ yields
\begin{equation}
g^{\beta \nu} c^{\alpha} - g^{\nu \alpha } c^{\beta} 
= -\left\langle {^0t}, \left( x^{\alpha} \partial^{\beta}
w - x^{\beta} \partial^{\alpha} w \right) x^{\nu}
\right\rangle. \label{ab35}
\end{equation}
Contracting finally with $g_{\nu \beta}$ on both sides yields
\begin{equation}
c^{\alpha} = \frac{1}{3} \left\langle {^0t}, x^2 \partial ^{\alpha} 
w -  x^{\alpha} x^{\beta}
\partial _{\beta} w \right\rangle.
\end{equation}
\item $|\alpha|=2$. We obtain from (\ref{stern1})
\begin{equation}
\left\langle {^0t},  \left( w- w_{\Lambda} \right)x^{\alpha_1}
x^{\alpha_2}\right\rangle = - \left( {D(\Lambda) ^{\alpha_1}}_{\beta_1} 
{D(\Lambda) ^{\alpha_2}}_{\beta_2} - \delta  ^{\alpha_1}_{\beta_1} 
 \delta  ^{\alpha_2}_{\beta_2} \right) c^{\beta_1 \beta_2}. 
\end{equation}
With (\ref{abcd}) and (\ref{abc}) we get
\begin{equation}
\left\langle {^0t},  
{(l^{\alpha \beta})^{\rho}}_{\mu} x^{\mu} \partial_{\rho} w
x^{\alpha_1}x^{\alpha_2}\right\rangle
= {(l^{\alpha \beta})^{\alpha_1}}_{\sigma} c^{\sigma \alpha_2}
+ {(l^{\alpha \beta})^{\alpha_2}}_{\sigma} c^{\sigma \alpha_1}.
\end{equation}
Inserting the form (\ref{form}) of the $(l^{\alpha \beta})$ and
contracting both sides with $g_{\beta \alpha_1}$ yields
\begin{equation}
4 c^{\alpha \alpha _2} - g^{\alpha \alpha _2} {c^{\mu}}_{\mu}
= - \left\langle {^0t}, x^{\alpha}x^{\alpha _2} x^{\sigma} 
\partial _{\sigma} w
- x^{\alpha _2} x^2 \partial^{\alpha} w \right\rangle.
\end{equation}
As in the case $|\alpha|=0$ we choose ${c^{\mu}}_{\mu}=0$ and obtain
(by setting $\alpha = \alpha _1$):
\begin{equation}
c^{(\alpha _1 \alpha_2)} = - \frac{1}{4} 
 \left\langle {^0t}, x^{\alpha _1}x^{\alpha _2} x^{\sigma} 
\partial _{\sigma} w
- x^2 x^{(\alpha _1}  \partial^{\alpha _2)} w \right\rangle.
\end{equation}
\end{enumerate}
We now assume that (\ref{result}) holds for all integers smaller than
$n$ and describe the induction $|\alpha|=n-2 \to |\alpha|=n$.  With
the examples of the beginning of the induction, it is easy to see how
the calculation is done for higher $|\alpha|$.  For $|\alpha|=n$ we
obtain instead of (\ref{ab35}) the following equation:
\begin{multline}
\sum_{i=1}^n \left(
g^{\beta \alpha _i} \delta^{\alpha}_{\mu} 
- g^{\alpha \alpha _i} \delta ^{\beta}_{\mu} \right)
c^{\mu \alpha _1 \ldots \check{\alpha}_i \ldots \alpha _n}= \\
= -\left\langle {^0t}, \left( x^{\alpha} \partial ^{\beta}
w - x^{\beta} \partial ^{\alpha}
w \right) x^{\alpha _1}x^{\alpha _2}
\ldots  x^{\alpha _n}\right\rangle.  \label{dfg}
\end{multline}
On the left hand side only the symmetric traceless part of the
$c^{\alpha}$ yields a contribution.  Contracting (\ref{dfg}) with
$g_{\beta \alpha _1}$ yields
\begin{multline}
(n+2) c^{\alpha \alpha _2 \ldots 
\ldots \alpha _n}
- \sum_{i \geq 2} g^{\alpha \alpha _i} 
{c_{\rho}}^{\rho \alpha_2 \ldots \check{\alpha}_i\ldots\alpha _n} = \\
=   -\left\langle t, \left( x^{\alpha} x^{\rho}
\partial_{\rho} w - x^2 \partial^{\alpha} w \right)
x^{\alpha _2} \ldots x^{ \alpha _n}\right\rangle. 
\label{2term}
\end{multline}
The second term on the left hand side of (\ref{2term}) is determined
by the induction hypothesis, because it is the solution of the problem
for $|\alpha|=n-2$ for the distribution $x^2 t$.  Setting $\alpha =
\alpha _1$ and symmetrizing in the indices $ \alpha_1 \ldots
\alpha_n$, the form (\ref{result}) is obtained.

\section{Dependence on $w$}
If we had chosen another testfunction $ w\in\Dcal(\Rv) $, our result
should differ only by Lorentz invariant counterterms. To use the
functional derivation, we notice, that the difference of two
admissible auxiliary functions has to be in $\Dcal_\omega(\Rv)$.
\begin{Def}
The functional derivation is defined by
\begin{gather*}
  \scp{\frac{\delta}{\delta w}F(w)}{\psi}:=
 \left. \frac{\dif}{\dif\lambda}F(w+\lambda\psi) \right\vert_{\lambda =0},
\end{gather*}
for all $\psi\in\Dcal_\omega(\Rv)$.
\end{Def}
Applying that definition to the first term of \eqref{fort}, we get:
\begin{gather}
  \scp{\frac{\delta}{\delta w}\scp{^0t}{W(\omega,w)\phi}}{\psi}
    =-\sum_{|\alpha|\leq\omega}
    \scp{^0t}{x^\alpha\psi}\frac{D^\alpha\phi(0)}{\alpha!}.
\label{eq:dwtR}
\end{gather}
Now we determine the contribution from $c$:
\begin{multline}
  \scp{\frac{\delta}{\delta w}c^{\alpha_{1}\cdots\alpha_{n}}(w)}{\psi}= \\
\begin{split}
  =&\frac{1}{n+2} \Bigl\langle \left( \D_\sigma x^\sigma
    x^{(\alpha_{1}}\cdots x^{\alpha_{n})}
    -\D^{(\alpha_1}x^{\alpha_2}\cdots x^{\alpha_{n})}x^2\right) {^0t}+ \\
  &\quad+\frac{n-1}{n}\Bigl( \D_\sigma x^\sigma
  x^{(\alpha_{1}}\cdots x^{\alpha_{n-2}}\eta^{\alpha_{n-1}\alpha_{n})}x^{2}+ \\
  &\qquad\qquad\qquad-\D^{(\alpha_1}x^{\alpha_2}\cdots
  x^{\alpha_{n-2}}\eta^{\alpha_{n-1}\alpha_n)}(x^{2})^2\Bigr){^0t}+ \\
  &\quad+\vdots \\
  &\quad+
\begin{cases}
  \dfrac{(n-1)\cdots2}{n\cdots3} &\Bigl(\D_\sigma x^\sigma
  x^{(\alpha_{1}}\eta^{\alpha_{2}\alpha_{3}}\cdots\eta^{\alpha_{n-1}\alpha_{n})}
  (x^{2})^{\frac{n-1}{2}}+ \\
  &\qquad\quad-\D^{(\alpha_1}\eta^{\alpha_2\alpha_3}\cdots\eta^{\alpha_{n-1}\alpha_n)}
  (x^{2})^{\frac{n+1}{2}} \Bigr){^0t},\psi\Bigr\rangle
  \ n \text{ odd}, \\
  \dfrac{(n-1)\cdots3}{n\cdots4} &\Bigl(\D_\sigma x^\sigma
  x^{(\alpha_1}x^{\alpha_2}\eta^{\alpha_{3}\alpha_{4}}\cdots\eta^{\alpha_{n-1}\alpha_{n})}
  (x^{2})^{\frac{n-2}{2}}+ \\
  &\qquad\quad-\D^{(\alpha_1}x^{\alpha_2}\eta^{\alpha_3\alpha_4}
  \cdots\eta^{\alpha_{n-1}\alpha_{n})} (x^{2})^{\frac{n}{2}}
  \Bigr){^0t},\psi\Bigr\rangle \ n \text{ even}
\end{cases} 
\end{split}
\label{eq:dMa}
\end{multline}
Since the singular order of every term is $\omega-n$ we are allowed to
differentiate strongly. The first line reads
\begin{multline}
  \D_\sigma x^\sigma x^{(\alpha_{1}}\cdots x^{\alpha_{n})}{^0t}
  -\D^{(\alpha_1}x^{\alpha_2}\cdots x^{\alpha_{n})}x^2\,{^0t} \\
  =(n+2)x^{(\alpha_{1}}\cdots x^{\alpha_{n})}{^0t}
  -(n-1)x^2\eta^{(\alpha_1\alpha_2}x^{\alpha_3}\cdots x^{\alpha_n)}{^0t} \\
  +x^{(\alpha_{1}}\cdots x^{\alpha_{n})}x^\sigma\D_\sigma{^0t}
  -x^2x^{(\alpha_{1}}\cdots x^{\alpha_{n-1}}\D^{\alpha_n)}{^0t}.
\label{eq:firstdMa}
\end{multline}
The Lorentz invariance condition
$\scp{{^0t}}{x^{[\alpha}\D^{\beta]}\psi} = 0,\ 
\forall\psi\in\Dcal^{\omega}(\Rd)$ allows to write $x^2\D^{\alpha_n}{^0t}
= x_\sigma x^\sigma\D^{\alpha_n}\,{^0t} = x_\sigma
x^{\alpha_n}\D^\sigma\,{^0t}$, so the last line of (\ref{eq:firstdMa})
vanishes. The second line of (\ref{eq:dMa}) is proportional to
\begin{multline}
  \D_\sigma x^\sigma x^{(\alpha_{1}}\cdots
  x^{\alpha_{n-2}}\eta^{\alpha_{n-1}\alpha_{n})}x^{2}\,{^0t}
  -\D^{(\alpha_1}x^{\alpha_2}\cdots
  x^{\alpha_{n-2}}\eta^{\alpha_{n-1}\alpha_n)}(x^{2})^2\,{^0t} \\
  =nx^2\eta^{(\alpha_1\alpha_2}x^{\alpha_3}\cdots x^{\alpha_n)}{^0t}
  -(n-3)(x^{2})^2 \eta^{(\alpha_1\alpha_2}\eta^{\alpha_3\alpha_4}x^{\alpha_5}\cdots
  x^{\alpha_n)}{^0t},
\label{eq:seconddMa}
\end{multline}
again two terms vanish because of Lorentz invariance of ${^0t}$.
Putting the $\frac{n-1}{n}$ in front of (\ref{eq:seconddMa}) and
adding to (\ref{eq:firstdMa}) we get $(n+2)x^{(\alpha_{1}}\cdots
x^{\alpha_{n})}\,{^0t} + \frac{n-1}{n}$ times the second term from
(\ref{eq:seconddMa}). This term cancels against the next line from
(\ref{eq:dMa}), and so on. For $n$ odd, the last line becomes
$\frac{(n-1)\cdots2}{n\cdots5} (x^2)^\frac{n-1}{2}
x^{(\alpha_1}\eta^{\alpha_2\alpha_3}\cdots \eta^{\alpha_{n-1}\alpha_{n})}{^0t}$,
which again cancels the line before. For $n$ even the last line is
\[
\frac{(n-1)\cdots3}{n\cdots4}\left( 4(x^2)^{\frac{n-2}{2}}
  x^{(\alpha_{1}}x^{\alpha_{2}}\eta^{\alpha_3\alpha_4}\cdots\eta^{\alpha_{n-1}\alpha_n)}
  -(x^2)^{\frac{n}{2}}\eta^{(\alpha_1\alpha_2}\cdots\eta^{\alpha_{n-1}\alpha_n)}\right)
{^0t},
\]
where again the first term is canceled by the line above but the
second gives a nontrivial contribution. We end up with
\begin{multline*}
  \scp{\frac{\delta}{\delta w}c^{\alpha_{1}\cdots\alpha_{n}}(w)}{\psi}=
  +\scp{^0t}{x^{\alpha_{1}}\cdots x^{\alpha_{n}}\psi}+ \\
  -\begin{cases}
    0,&n\text{ odd}, \\
    \frac{2(n-1)!!}{(n+2)!!}  \scp{^0t}{(x^2)^\frac{n}{2}\psi}
    \eta^{(\alpha_1\alpha_2}\cdots\eta^{\alpha_{n-1}\alpha_n)}, 
    &n\text{ even.}
\end{cases}
\end{multline*}
Using this result and \eqref{eq:dwtR} we find:
\begin{align*}
  \scp{\frac{\delta}{\delta w}\scp{t}{\phi}}{\psi}
  &=-\sum_{\substack{n=0\\ n\text{ even}}}^{\omega}\frac{d_{n}}{n!}
  \square^{\frac{n}{2}}\phi(0), \\
  d_{n}&:= \frac{2(n-1)!!}{(n+2)!!}\scp{^0t}{(x^{2})^{\frac{n}{2}}\psi}, 
\end{align*}
where we set $ d_{0}=1 $.

\section{Conclusion and Outlook}
We give the explicit form of Lorentz invariant Epstein-Glaser
renormalized distributions in one argument.  In case of distributions
depending on more than one argument the same calculation yields the
totally symmetric coefficients $c$. But then the other coefficients do
not vanish in general.  In $\varphi^4$-theory this is already
sufficient for the renormalization up to third order.  The general
case for Lorentz invariant and Lorentz covariant distributions is
treated in \cite{prep:prange2}.

\section*{Acknowledgements} We thank Prof.\,K.\,Fredenhagen for useful hints
and Michael D\"utsch for reading the manuscript.


\begin{thebibliography}{99}

\bibitem{stue} E.C.G.Stueckelberg and D. Rivier, Helv. Phys. Acta,
{\bf 22}(1949) 215.
E.C.G.Stueckelberg and J. Green, Helv. Phys. Acta,
{\bf 24}(1951) 153.

\bibitem{meister} N.N. Bogoliubov and D. Shirkov, {\it Introduction
to the theory of quantized fields}, John Wiley and Sons, 1976, 3rd edition.

\bibitem{epstein} H. Epstein and V. Glaser, {\it The role of locality
in perturbation theory}, Ann. Inst. Henry Poincar\'e -Section A,
Vol. XIX, n.3 (1973)p.211.

\bibitem{stora} G. Popineau and R. Stora, \ {\it A Pedagogical Remark
on the Main Theorem of Perturbative Renormalization Theory},
unpublished preprint.

\bibitem{scharf} G. Scharf, {\it Finite Quantum Electrodynamics:
The Causal Approach}, Springer-Verlag, 1995, 2nd edition.

\bibitem{fred} K. Fredenhagen, 
{\it Renormierung auf Gekruemmten Raumzeiten}, lecture notes

R. Brunetti and K. Fredenhagen, {\it Interacting 
Quantum Fields in Curved Space: Renormalizability of $\phi ^4$},
gr-qc/9701048, {\it Proceedings of the Conference 'Operator Algebras 
and Quantum Field Theory'}, held at Accademia Nazionale dei Lincei, Rome,
July 1996. 

R. Brunetti and K. Fredenhagen, {\it Microlocal
Analysis and Interacting Quantum Field Theories: Renormalization
on Physical Backgrounds}, math-ph/9903028. 

\bibitem{prep:prange2}D. Prange, {\it Lorentz Covariance in 
Epstein-Glaser Renormalization}, hep-th/9904136.


\end{thebibliography}
\end{document}